# Room Temperature Coherent Control of Spin Defects in hexagonal Boron Nitride

Andreas Gottscholl[1], Matthias Diez[1], Victor Soltamov[1], Christian Kasper[1], Andreas Sperlich[1], Mehran Kianinia[2], Carlo Bradac[3], Igor Aharonovich[2,4], Vladimir Dyakonov[1,*]

[1] Experimental Physics 6 and Würzburg-Dresden Cluster of Excellence ct.qmat, Julius Maximilian University of Würzburg, 97074 Würzburg, Germany
[2] School of Mathematical and Physical Sciences, University of Technology Sydney, Ultimo, NSW 2007, Australia
[3] Department of Physics & Astronomy, Trent University, 1600 West Bank Dr., Peterborough 28 ON, K9J 0G2, Canada
[4] Centre of Excellence for Transformative Meta-Optical Systems, University of Technology Sydney, Ultimo, NSW 2007, Australia

*Optically active defects in solids with accessible spin states are promising candidates for solid state quantum information and sensing applications. To employ these defects as quantum building blocks, coherent manipulation of their spin state is required. Here we realize coherent control of ensembles of boron vacancy ($V_B^-$) centers in hexagonal boron nitride (hBN). Specifically, by applying pulsed spin resonance protocols, we measure spin-lattice relaxation time ($T_1$) of 18 μs and spin coherence time ($T_2$) of 2 μs at room temperature. The spin-lattice relaxation time increases by three orders of magnitude at cryogenic temperature. Furthermore, employing a two- and three-pulse electron spin-echo envelope modulation (ESEEM) we separate the quadrupole and hyperfine interactions with the surrounding nuclei. Finally, by applying a method to decouple the spin state from its inhomogeneous nuclear environment – a "hole-burning" – the spectral optically detected magnetic resonance linewidth is significantly reduced to several tens of kHz, thus extending the spin coherence time by a factor of three. Our results are important for employment of van der Waals materials for quantum technologies, specifically in the context of using hBN as a high-resolution quantum sensor for hybrid quantum systems including 2D heterostructures, nanoscale devices and emerging atomically thin magnets.*

Van der Waals (vdW) crystals have recently emerged as a promising family of materials to investigate light matter interaction at the nanoscale[1-4]. Out of a growing suite of vdW crystals, hexagonal boron nitride (hBN) stands out owing to its ability to host optically active defects that emit single photons of light, whilst displaying spin-optical quantum properties at room temperature[5,6]. A specific defect of interest is the negatively charged boron vacancy ($V_B^-$) center. This atom-like defect possesses a spin triplet ground state (S=1), which can be prepared, manipulated and optically read out through a combination of microwave and laser excitation cycles[6]; a feat that makes it particularly appealing for quantum sensing and spintronic applications. In the realm of vdW heterostructures, having an optically active spin sensor confined in an intrinsically atom-thin, two-dimensional host is highly desirable, as it allows for achieving high resolution sensing strategies with specific requirements such as, for instance, imaging the magnetic domains in 2D ferromagnets[7-9] or magnetic superlattices[10].

However, before these defects can be utilized in practical implementations, full understanding and, ultimately, control of their coherent properties must be acquired. In this work we demonstrate coherent control of an ensemble of $V_B^-$ defects in hBN. While coherent control of spin qubits in 3D crystals (e.g. diamond, silicon carbide or rare earth ions in glass)[11-15] has been demonstrated, spin defects in vdW crystals are still unexplored. We further show that the coherence properties of $V_B^-$ spin ensembles are influenced by the coupling with the surrounding nuclei spin bath ($^{14}$N with I=1, $^{10}$B, I=3 and $^{11}$B, I=3/2). By applying optically detected two and three-pulse electron spin-echo envelope modulation (ESEEM), we were able to probe magnetic and quadrupole fields of surrounding nuclei and obtain information on hyperfine coupling and quadrupole splitting – although an exact assignment to nitrogen or boron in their corresponding coordination shells remains ambiguous. Remarkably, by implementing a two-frequency optically detected magnetic resonance (ODMR) technique to the $V_B^-$ spin system, we demonstrate the ability to selectively saturate one of the hyperfine transitions and effectively decouple the electron spin system from the nuclear bath, which allows estimating the upper limit of the highest achievable coherence times.

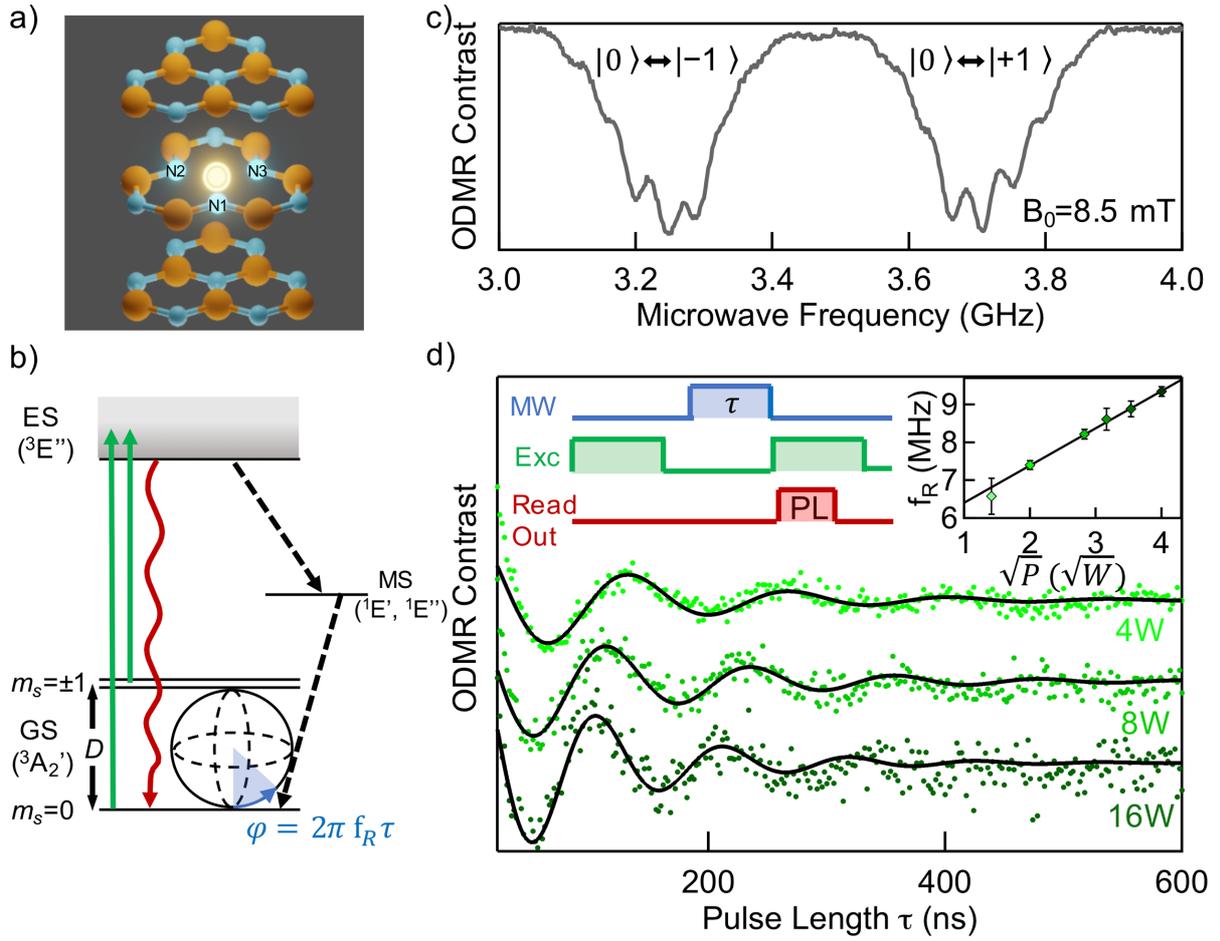

**Figure 1.** Coherent manipulation of $V_B^-$ spin centers in hBN at room temperature. a) Schematic of the $V_B^-$ defect in hBN: a negatively charged boron vacancy (bright spot) surrounded by three equivalent nitrogen atoms (blue spheres, N1-N3). b) $V_B^-$ defect energy diagram and the optical pumping cycle comprising excitation to excited state (ES) (green), radiative recombination (red) and non-radiative decay to ground state (GS) via metastable state (MS) (dashed lines). D denotes ZFS. The $|0\rangle \leftrightarrow |-1\rangle$ transition is represented on a Bloch sphere. The GS $|0\rangle$ state is spin polarized by a laser (green) and can be manipulated to an angle $\varphi$ after applying microwaves with the Rabi frequency $f_R$ for a time $\tau$. c) CW ODMR spectrum with two characteristic transitions in a magnetic field of 8.5 mT. Both signals reveal structure due to hyperfine interaction with three next neighboring nitrogen atoms d) Optically detected Rabi oscillations on transition $|0\rangle \leftrightarrow |-1\rangle$ after initialization into $|0\rangle$ state for different microwave power. The inset schematically shows the pulse sequence, in which the first excitation laser pulse (green) is used for the initialization of the spin state and the second one for a readout (red) of the current state after the microwave pulse (blue). Rabi frequency $f_R$ versus square root of microwave power P, which is proportional to $B_1$ field, is also shown.

A schematic of the $V_B^-$ defect is shown in Figure 1a. The defect consists of a missing boron atom in the hBN crystal in a negatively charge state. The corresponding simplified energy level diagram is shown in Fig. 1b. The defect has a nominal $D_{3h}$ symmetry (lower symmetries are expected due to strain and reorganization) and the main optical transition takes place between the $^3E''$ and $^3A'_2$ levels. When excited with a 532-nm laser source, the defect emits at a wavelength centered ~ 850 nm. Inset shows the evolution of the $|0\rangle$ sub-state on a Bloch sphere under the action of the microwave field.

Fig. 1c shows the CW-ODMR spectrum of the $V_B^-$ defects at room temperature. The signal is centered around the 3.5 GHz frequency, corresponding to the zero-field splitting (ZFS). The photoluminescence (PL) intensity signal is frequency-dependent. The two distinct features at frequencies $\nu_1$ and $\nu_2$ correspond to the ground state spin transitions $|0\rangle \leftrightarrow |-1\rangle$ and $|0\rangle \leftrightarrow |+1\rangle$, and occur because the ground $m_s = \pm 1$ states scatter less photons than the $m_s = 0$ state, through the excited state. Both features show a clearly resolved hyperfine splitting with seven peaks due to three equivalent nitrogen nuclei (2nI+1=7, with n=3, I=1) in the hBN plane surrounding the missing boron atom. To explain the observed transitions and their variation with the magnetic field, we use the spin-Hamiltonian, which describes electron spin-spin (ZFS), electron and nuclear Zeeman, electron-nuclear hyperfine (HFI) and electric quadrupole (QI) interactions:

$$H = \underbrace{D(S_z^2 - S(S+1)/3) + E(S_x^2 - S_y^2)}_{ZFS} + \underbrace{g_e \mu_B \boldsymbol{BS}}_{e-Zeeman} + \underbrace{\boldsymbol{SAI}}_{HFI} + \underbrace{Q(I_z^2 - I(I+1)/3)}_{QI} - \underbrace{g_N \mu_N \boldsymbol{BI}}_{n-Zeeman}, \quad (1)$$

where D and E are the ZFS parameters, S is the total electron spin (S = 1 for $V_B^-$), g is the Landé factor, $\mu_B$ is the Bohr magneton, $\mu_N$ is the nuclear magneton, B is the static magnetic field, $S_{x,y,z}$ are the spin-1 operators, **A** is the HFI tensor, I and $I_z$ are nuclear operators, Q is the quadrupole coupling constant. The hyperfine coupling constant due to coupling with $^{14}$N (I = 1) in the first coordination shell A = 47 MHz is known from the experiment and confirmed by calculations[6,16].

To get access to the spin dynamics of the $V_B^-$, and to determine their spin relaxation times $T_1$ and $T_2$, we perform pulsed ODMR measurements. They are based on preparing the ground spin state of the system by optical excitation, applying microwave pulses of variable length to coherently manipulate the spin state, and lastly performing an optical readout of the state. Fig. 1d shows Rabi-oscillations with a characteristic frequency $f_R$ in the megahertz range, which depends on the microwave power and hence the strength of the $B_1$ component of the microwave field, as shown in the inset. Importantly, these measurements allow to calibrate the microwave pulse length at a given power for subsequent pulsed ODMR experiments.

The upper limit for spin coherence is given by the spin-lattice relaxation time $T_1$, which we determine with a standard π-pulse sequence[11,12], as shown in Figure 2a.

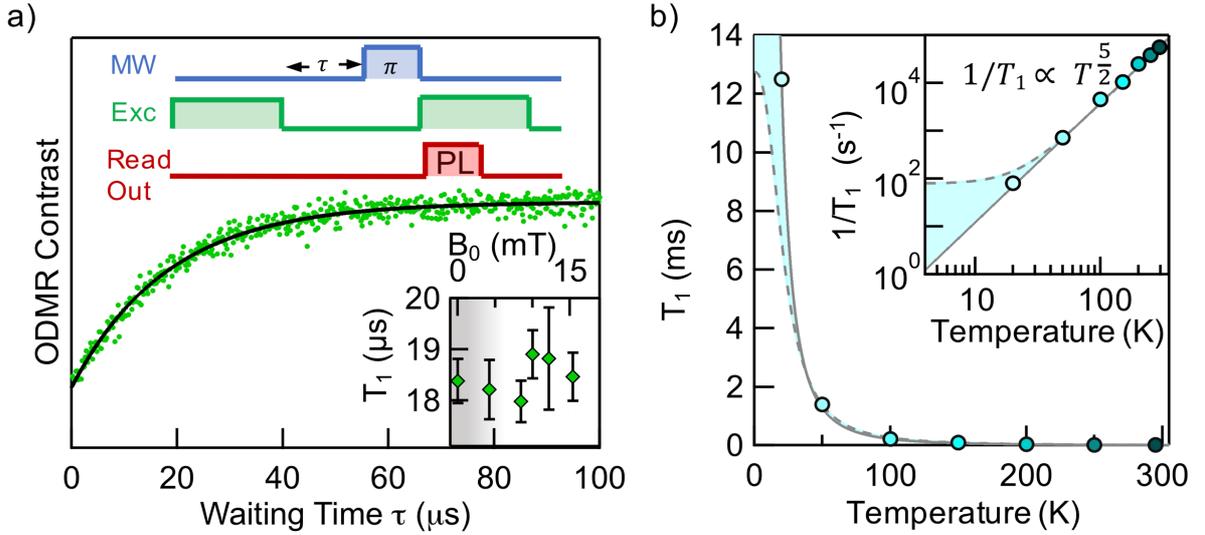

*Figure 2.* Spin-lattice relaxation dynamics of $V_B^-$ spin centers in hBN. a) ODMR signal decay with waiting time τ at T = 300K. The pulse sequence is schematically shown, in which the duration of microwave π-pulse is known from Rabi-experiments and the waiting time τ is varied. Inset schematically shows $T_1$ versus external magnetic field (0-15 mT). The grey background indicates the magnetic field range where two ODMR transitions $v_1$ and $v_2$ overlap. b) Temperature dependence of the spin-lattice relaxation time of $V_B^-$ spin centers in hBN. The $T_1$ time increases by three orders of magnitude from 18 μs to 12.5 ms, when lowering the temperature to 20 K. Inset: a log-log plot for the spin-lattice relaxation rate ($1/T_1$) shows a $T^{5/2}$ behavior over a broad temperature range. An upper boundary for the fastest spin-lattice relaxation rate can be estimated by assuming a saturation of the $T_1$ time at 20K (highlighted in blue).

The $T_1$ time is around 18 μs and does not seem to be influenced by an external magnetic field, even if the two ODMR peaks collapse and form a single peak at B = 0 (see inset to Fig. 2a). The grey background indicates the overlap of the two transitions $v_1$ and $v_2$ below 5 mT. This can be explained by the large ZFS compared to the small external magnetic field applied. Note that ZFS of $V_B^-$ centers in hBN is 3.5 GHz and the magnetic fields applied in our experiments are not large enough to induce the level-anticrossing (B ≈ 125 mT). Hence, the local magnetic field at the spin defect site dominates the spin properties. This leads to a robust spin-lattice relaxation dynamic of the $V_B^-$ defect, independent of external magnetic perturbation fields.

To gain further insights into the nature of spin-lattice relaxation mechanisms, e.g. the interaction with lattice phonons, we probe the temperature dependence of $T_1$. In this case one would vary the spectrum of vibronic lattice modes, which are an effective relaxation channel if they have energies comparable to the Larmor frequency of the electron spin. The $V_B^-$ ZFS is very large, so one can expect a relaxation behavior similar to the nitrogen vacancy (NV) centers in diamond, where the direct one-phonon absorption and emission processes are neglected at higher temperatures[17]. Nevertheless, the 2D character of the hBN and the symmetry of the spin defects are expected to influence their vibronic properties compared to 3D crystals[18,19], although detailed calculations about these effects in the presence of spin defect or experimental data are not yet available.

To study the behavior of spin-phonon contributions to $T_1$, we use the pulse sequence shown in Fig. 2a and varied the temperature between 300 K and 20 K. Fig. 2b reveals a monotonic growth of $T_1$, as the temperature decreases, up to a value of 12.5 ms at T = 20 K. In general, the $T_1$ time is governed by the following processes[20,21]:

$$\frac{1}{T_1} = A_0 + A_1 T + A_s T^s + \frac{R}{\exp\left(\frac{\Delta}{k_B T}\right) - 1} \qquad (2)$$

Single-phonon scattering processes are considered by the linear term $A_1$. Two-phonon processes (direct transition, Stokes, anti-Stokes and spontaneous emission) are described by the $A_s$ term (higher orders s+1 and s+2 are neglected), where s=4ν+2d-3[21]. Since d is the dimension of the system and ν is a phenomenological spin-phonon coupling factor reflecting the symmetry of the lattice (ν=1/2 for a non-cubic), one expects s=5 for the NV centers in diamond and silicon vacancies ($V_{Si}$) in silicon carbide (SiC). For a non-cubic 2D system such as hBN (d=2, ν=1/2) one expects s=3. Orbach-type processes resulting from quasi-localized phonon modes are described by an exponential contribution with the characteristic energy $\Delta = \hbar\omega_{loc}$. Therefore, the spin-lattice relaxation rate is ultimately limited by $A_0$ which provides the longest achievable $T_1$ time. In order to quantify the observed temperature dependence shown in Fig. 2b, a double logarithmic scaling of the spin-lattice relaxation rate $1/T_1$ is depicted in the inset. All observed values can be fitted with only one line with a slope of s=5/2 which corresponds to $\frac{1}{T_1} \sim T^{5/2}$. This value differs from the expected values of s=3 (s=5) for a 2D (3D) system mentioned above. Since the experimental values are all in the linear regime (in the log-log scale), it is not possible to make any further assumptions regarding other spin-lattice rate contributions such as linear terms or Orbach-related effects. Nevertheless, an estimate for the lowest relaxation rate corresponding to the longest measured $T_1$ = 12.5 ms can be given. Assuming that the spin-lattice relaxation rate starts to level off and becomes temperature independent at temperatures below 20 K[12,17,20], an upper limit for the term $A_0 < 80$ s$^{-1}$ can be set, as highlighted in blue in Fig. 2b.

After we have determined the $T_1$ relaxation time, we proceed to measure the $T_2$ time of the $V_B^-$ defects employing a Hahn spin-echo sequence[22-24]. The pulse sequence applied is π/2-τ-π-τ-*echo* and the results are shown in Fig. 3a. Note that to enable optical detection of spin echo, an additional π/2-pulse[25-27] is required after the standard spin-echo pulse sequence to read out the spin polarization of a well-defined state (bright state $|0\rangle$) vs. dark states $|\pm 1\rangle$). The ODMR transient shows the exponential rise from which we determine the spin coherence $T_2$ time. A room temperature spin-spin coherence time of $T_2 \approx 2$ μs can be extracted for an external magnetic field of 8.5 mT. As shown in Fig. 3b, $T_2$ increases slightly to about 3 μs by increasing magnetic field above the region where the ODMR signals overlap (grey zone). Furthermore, the ODMR transient is superimposed by oscillations containing two frequencies, as determined by Fourier analysis and plotted for different magnetic fields in Fig. 3c. The effect is known as electron spin-echo envelope modulation (ESEEM)[28] and is widely used to study interactions between electron and nuclear spins. In ESEEM, the electron spin-echo envelope exhibits amplitude modulation which corresponds to the NMR frequencies of nuclei, which are coupled to the electron spin.

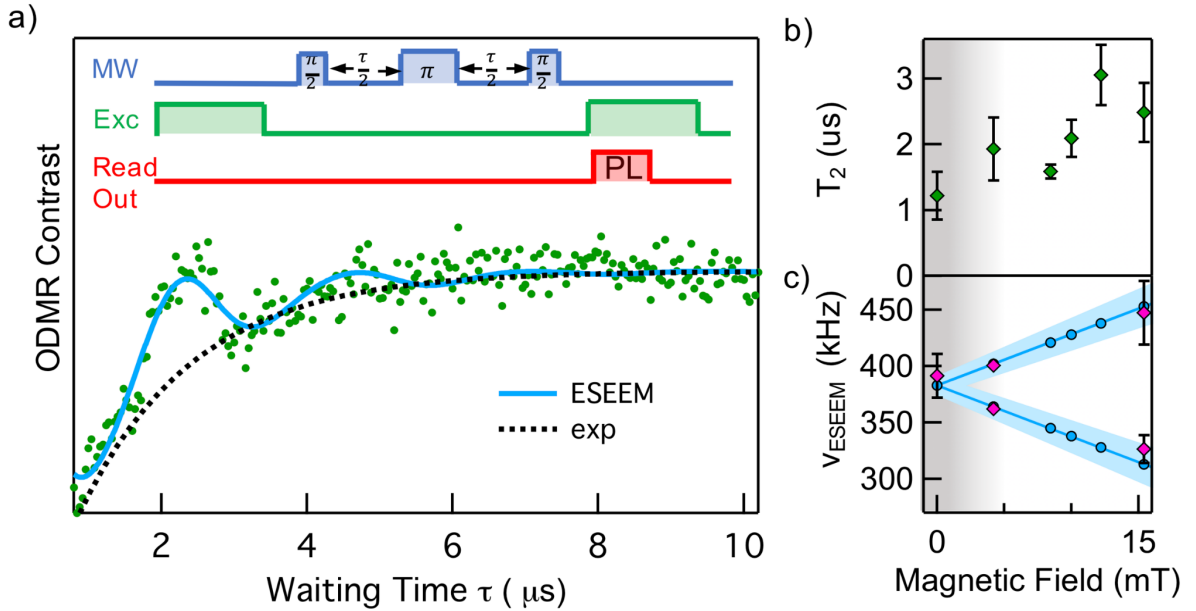

*Figure 3. Spin-spin relaxation dynamics and nuclear modulation effects at room temperature. a) The applied spin-echo pulsed sequence including the projection pulse π/2 and the ODMR contrast vs. waiting time τ (green) are shown. The expected exponential rise (black dotted line) is superimposed by the oscillations (green and blue) due to coupling of electron spins to nuclei. (ESEEM effect, see text). The fit with modulated exponential function (blue) allows to estimate the ESEEM frequencies and the $T_2$ time. b) $T_2$ time versus magnetic field. $T_2$ slightly decreases in the magnetic field range where ODMR transitions $\nu_1$ and $\nu_2$ overlap (grey area), indicating the presence of additional decoherence, e.g. due to hyperfine interaction. c) ESEEM frequency versus external magnetic field. Plotted are frequencies obtained from*

For the $V_B^-$ only one ESEEM frequency is expected if the hyperfine interaction of the electron spin with three equivalent nitrogen nuclei (I=1) in the first coordination sphere is considered. For all nuclei with I>1/2, however, the quadrupole term becomes important, as described in Eq. (1), since it modifies the NMR frequency spectrum considerably. In this case, one expects six frequencies derived from the combination of the Zeeman nuclear and the quadrupole splitting of the surrounding nuclei, as shown in Supporting Figure S1. Note that additional splitting (with corresponding frequencies) are expected when the $^{10}B$ (I=3) nuclei are taken into account. Due to experimental limitations, we can only resolve two of them. Fig. 3c plots a global fit over all $T_2$ transients measured at different magnetic fields (blue circles). Additional frequency values for other magnetic fields obtained by three-pulse ESEEM (see Supporting Figure S2) are also shown (pink diamonds) and they complement the field dependence obtained by the two-pulse ESEEM within the error margin (shown in light blue). The ESEEM frequencies evolve linearly with the magnetic field B following the law $\nu_{ESEEM} = |\nu_0 \pm \gamma B|$, where $\gamma = 4.54 \pm 0.59\, kHz/mT$ and $\nu_0 = 382.9 \pm 4.4\, kHz$ is the frequency at B = 0. In the absence of magnetic field, the nuclear Zeeman splitting is zero and ESEEM arises only from the nuclear quadrupole and/or hyperfine interactions[29] and can be used to determine the quadrupole coupling as Q= $\nu_0$ = 383 kHz (Eq.1), as shown in Fig. 3c. The experimentally derived $\gamma$ is close to the tabulated value of the $^{10}B$ nuclear gyromagnetic ratio ($\gamma(^{10}B) = 4{,}575\, kHz/mT$), thus the observed oscillations are possibly due to the coherent coupling of the electron spin with the $^{10}B$ nuclear spin bath. Note that the coupling of the electron spin VB- with the neighboring $^{14}N$ nuclear spin with slightly smaller gamma ($\gamma(^{14}N) = 3.076\, kHz/mT$) cannot be fully disregarded since it is also within the error margins of the $\nu_{ESEEM}(B)$ dependence in the Fig. 3c.

A slight increase of the spin coherence time $T_2$ with increasing magnetic field (Fig. 3b) may be due to the partial suppression of hetero- or homonuclear spin pair flip-flop processes (cross-resonances) in a moderate magnetic field[30], however the nuclear spin bath of the hBN lattice with $^{14}N$ (99.6%) (nuclear spin I=1), $^{11}B$ (80.1%) and $^{10}B$ (19.9%) that have nuclear spin of 3/2 and 3, respectively, remains a major factor of decoherence. Observed ESEEM frequencies clearly show that the electron spin of $V_B^-$ is coherently coupled to a nuclear spin in the first and second coordination shells, but the exact decoherence mechanism is not yet understood.

To further explore the origins of the ODMR line broadening and hence the spin-dephasing, we apply a two-frequency CW ODMR technique, which eliminates the contributions of the magnetic nuclei. The so-called "hole-burning" technique[31-34] is realized by introducing a second frequency (pump) which is fixed within the inhomogeneously broadened ODMR line. In Fig. 4a, a standard CW ODMR spectrum is shown (black curve). Applying a second microwave frequency results in a sharp dip in a particular ODMR transition at the position of the applied frequency. This is shown for three selected resonances at 3.2 GHz (pink), 3.25 GHz (green) and 3.3 GHz (orange), respectively. The pump microwave field excites modes in certain spin packets, and if the power is sufficiently high, it saturates them, resulting in a reduction in ODMR contrast.

To evaluate the spin hole-burning and extract the "hole" spectrum directly, we zoom into a selected transition at 3.25 GHz, as shown in Fig. 3c. The spectral shape of the hole can be fitted by a gaussian function with full width at half maximum (FWHM) of 15.5 MHz with a sharp spike in the middle. The linewidth of a broad component is 15.5 MHz and much narrower than the ODMR signal with the linewidth of several hundreds of MHz due to hyperfine broadening. Thus, by saturating an individual hyperfine transition, we are able to eliminate the linewidth contribution from magnetic nuclei and separate a spin package. As shown in Fig. 4d, a 500 kHz frequency sweep around the center of the "hole" reveals a spike which is perfectly reproducible by two Lorentz functions with 84.7 kHz (blue) and 25.5 KHz (yellow) linewidths, respectively. This narrow spike can be explained by coherent population oscillations, where the ground state population of a two-level quantum system oscillates at the beat frequency between the pump and probe (swept frequency) fields[31,32,35]. Such oscillations can be observed only if the beat frequency is less than, or approximately equal to, the inverse of the population relaxation time.

Projecting the spectral hole-burning in optical spectroscopy to the $V_B^-$ spin system, the width of the spike can be related to the population decay rates (relevant spin-relaxation processes)[31,36]. Since there are two contributions to the linewidth we assign the narrow contribution (25.5 kHz) to the spin-lattice relaxation rate and extract a $T_1^{HB}$ of 25.0 μs. Consequently, the second contribution (84.7 kHz) is assigned to $T_2^{HB}$ of 7.5 μs. In particular, the spin-spin relaxation time estimated in this way exceeds the value measured directly by pulsed ODMR by a factor of three after the broadening contributions are strongly suppressed by the surrounding nuclear bath, and can therefore be regarded as the upper limit for the relaxation rates for $V_B^-$ in hBN at room temperature. Note, the spectral hole is mirrored relative to the ODMR spectrum symmetry axis at D/h=3.48 GHz and also appear in the second ODMR transition (~ 3.7 GHz range), as indicated by asterisks in Fig. 4a.

To understand this phenomenon, one can use a simplified energy diagram in which only the hyperfine interaction with three equivalent nitrogen nuclei is considered, see Fig. 4b. (Supplementary Figure S1 gives a detailed overview of all energy contributions.) As soon as the pump frequency has saturated the transition, e.g. between the $m_S$=0 state and the $|-1,-1\rangle$ sub-state of $m_S$= −1 state, the hole in the left (low frequency) ODMR signal is "burned". Hence, the state $m_S$=0 is already depleted and is kept so, therefore the intensity of the transition to the sub-state $|+1,-1\rangle$ of the state $m_S$= +1 (high frequency ODMR signal) will also be lower. This applies to all transitions that correspond to the nuclear spin quantum numbers $m_I$.

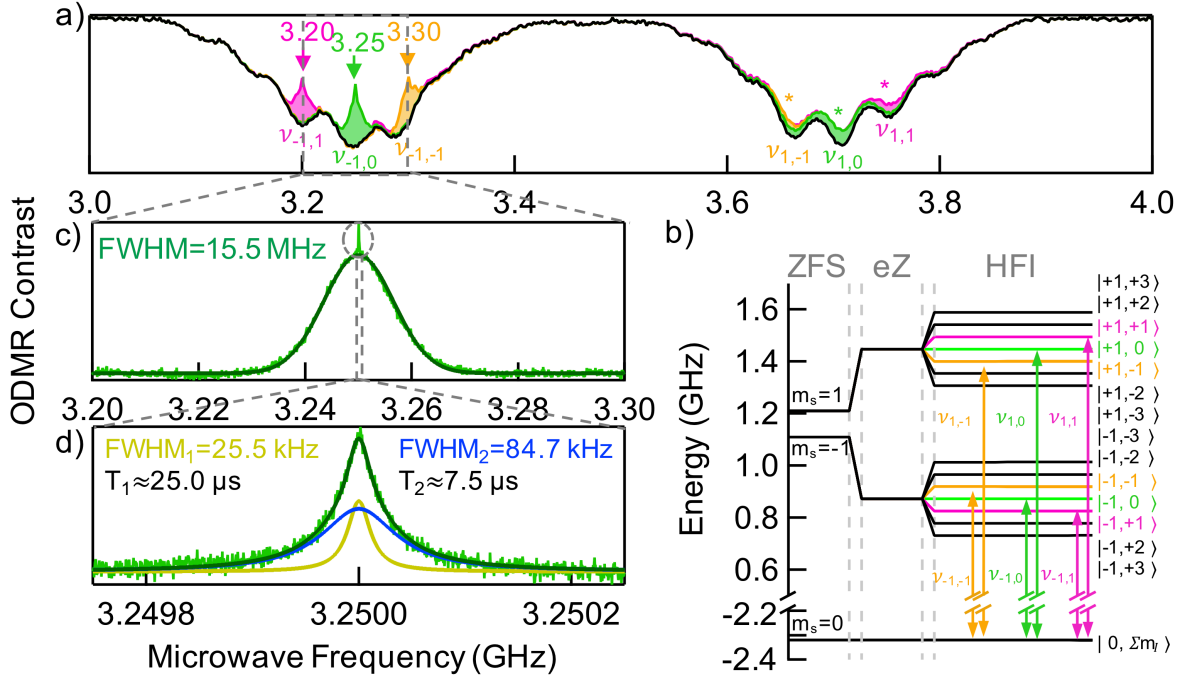

*Figure 4.* CW multi-frequency ODMR measurements at T=300K. a) A standard CW ODMR measurement is plotted in black. Applying a second microwave frequency (pump) leads to a dip ("hole") indicated by an arrow (3.20 GHz pink, 3.25 GHz green, 3.30 GHz orange). A second peak (indicated by an asterix) appears symmetrically to the center of the spectrum (3.48 GHz) for the transitions from $m_S$=0 to $|-1,-1\rangle$ and $|+1,-1\rangle$ sub-states, respectively. b) simplified energy diagram illustrating the hyperfine splitting due to three equivalent nitrogen nuclei with the respective electronic and nuclear spin quantum numbers $m_S$ and $m_I$. c) The hole-spectrum. By modulating the fixed (pump) frequency instead of the swept (probe) frequency the pristine hole-spectrum can be obtained directly. The signal consists of a broad inhomogeneously broadened Gaussian peak and a very narrow spike on top in the center. d) The narrow peak is due to coherent population oscillations and consists of two contributions and can be fitted with two Lorentz functions (yellow and blue, sum of both in green). The inset shows the respective FWHM and the inversed values assigned to spin-relaxation times $T_1$ and $T_2$.

**Conclusions**

To conclude, we demonstrated coherent control of the $V_B^-$ spin ensemble in hBN at room temperature. Our results suggest reasonably long spin-relaxation times $T_1 \approx 18$ µs and $T_2 \approx 2$ µs at room temperature that can be significantly increased to $T_1 \approx 12.5$ ms at cryogenic temperatures. The behavior is governed by spin-phonon interaction and follows a power law of $T^{-5/2}$. Employing a spin hole-burning technique, we determine the upper limit of the spin-relaxation times to $T_1^{HB} \approx 25.0$ µs and $T_2^{HB} \approx 7.5$ µs at room temperature by suppressing the inhomogeneous broadening due to the surrounding nuclear bath. Additionally, the electronic spin system can be further decoupled from the detrimental nuclear bath, as was demonstrated for other 3D systems[37-39], whereby the spin coherence times needs to be seen in relation to other properties depending on the applications[12,40]. The ability to engineer $V_B^-$ defects in thin hBN flakes on demand[41] offers promising perspectives for deploying quantum sensing with vdW crystals. For instance, positioning a monolayer or few layer hBN with embedded $V_B^-$ within a stack of 2D materials can provide an in-situ magnetometer to sense the surrounding electromagnetic fields[42,43] or couple the spin to the valley degree of freedom from the surrounding excitons in transition metal dichalcogenides[44].

## Methods:

**hBN Sample** Monocrystalline hBN were neutron irradiated in the Triga Mark I IPR-R1 nuclear reactor (CDTN, Brazil), with a thermal flux of $4 \cdot 10^{12}$ n cm$^{-2}$s$^{-1}$ for 16 h reaching an integrated dose of approximately $2.3 \cdot 10^{18}$ n cm$^{-2}$. For more details about sample preparation by fast neutrons and thermal stability of produced defects see reference[45].

**CW ODMR** All low field ODMR measurements were performed with a home-made confocal setup. A 532nm laser (Cobolt Samba 100) was coupled into a 50-μm fiber and focused onto the sample with a x10 objective (Olympus LMPLN10XIR) which excited a spot on the sample with a diameter of about 10 μm. The photoluminescence was separated from the laser by a dichroic mirror and the residual laser light was absorbed by a 532-nm long-pass filter. Afterwards the photoluminescence was coupled into a 600-μm fiber and guided to an avalanche photodiode (Thorlabs APD440A). A 0.5-mm-wide copper stripline was used to apply microwaves to the on top laying hBN sample. The microwaves were generated by a signal generator (Stanford Research Systems SG384) and amplified by a Mini Circuits ZVE-3W-83+ amplifier. A lock-in detection was applied (Signal Recovery 7230 lock-in amplifier) by using an on-off modulation of the microwaves. For an external magnetic field, a permanent magnet was mounted below the sample.

**CW hole-burning ODMR** Additionally to the described CW ODMR setup, a second identical microwave source is added to the first frequency with a combiner (Mini Circuits ZFRSC-183-S+) in front of the amplifier (Mini Circuits ZVE-3W-83+).

**Pulsed ODMR** A pulse blaster card (PulseBlasterESR PRO 500 MHz) was used for building the described pulse sequences. The optical excitation is pulsed by an AOM (AA.MT250-A0,2-VIS) and the microwave pulses are formed by a fast switching solid state microwave switch (ZASWA-2-50DR+) directly after the microwave generator. The amplification to the required microwave power is performed by a Vectawave VBA2060-25. For the detection, an APD with a higher bandwidth was used (Thorlabs APD120A/M) which is than digitized by a digitizer card (GaGe Compuscope).

## Data availability

All relevant data are available from the corresponding author upon request.

## Acknowledgements

A.G and V.D. acknowledge financial support from the DFG through the Würzburg-Dresden Cluster of Excellence on Complexity and Topology in Quantum Matter—ct.qmat (EXC 2147, project-id 39085490) and DY18/13-1. V.S. gratefully acknowledges the financial support of the Alexander von Humboldt (AvH) Foundation. The Australian Research council (via DP180100077, DP190101058 and CE200100010), the Asian Office of Aerospace Research and Development grant: FA9550-19-S-0003 are gratefully acknowledged. I.A. is grateful for the Humboldt Foundation for their generous support.

## Author contributions

The experimental set-ups were implemented and the CW, hole-burning and pulsed ODMR measurements were performed by A.G., M.D., C.K., V.S. and A.S.. I.A., M.K. C.B. assisted with sample fabrication. V.D. and I.A. conceived and supervised the project. All the authors contributed to analysis of the data, discussions and to the writing of the paper.

## Competing interests

The authors declare no competing interests.

## Additional information

Supplementary Information is available for this paper at https://...

# Room Temperature Coherent Control of Spin Defects in hexagonal Boron Nitride

Andreas Gottscholl[1], Matthias Diez[1], Victor Soltamov[1], Christian Kasper[1], Andreas Sperlich[1], Mehran Kianinia[2], Carlo Bradac[3], Igor Aharonovich[2,4], Vladimir Dyakonov[1,*]

[1] Experimental Physics 6 and Würzburg-Dresden Cluster of Excellence ct.qmat, Julius Maximilian University of Würzburg, 97074 Würzburg, Germany
[2] School of Mathematical and Physical Sciences, University of Technology Sydney, Ultimo, NSW 2007, Australia
[3] Department of Physics & Astronomy, Trent University, 1600 West Bank Dr., Peterborough 28 ON, K9J 0G2, Canada
[4] Centre of Excellence for Transformative Meta-Optical Systems, University of Technology Sydney, Ultimo, NSW 2007, Australia

*Energy level diagram displaying the hyperfine and quadrupole interaction with three nitrogen nuclei in the vicinity of boron vacancy*

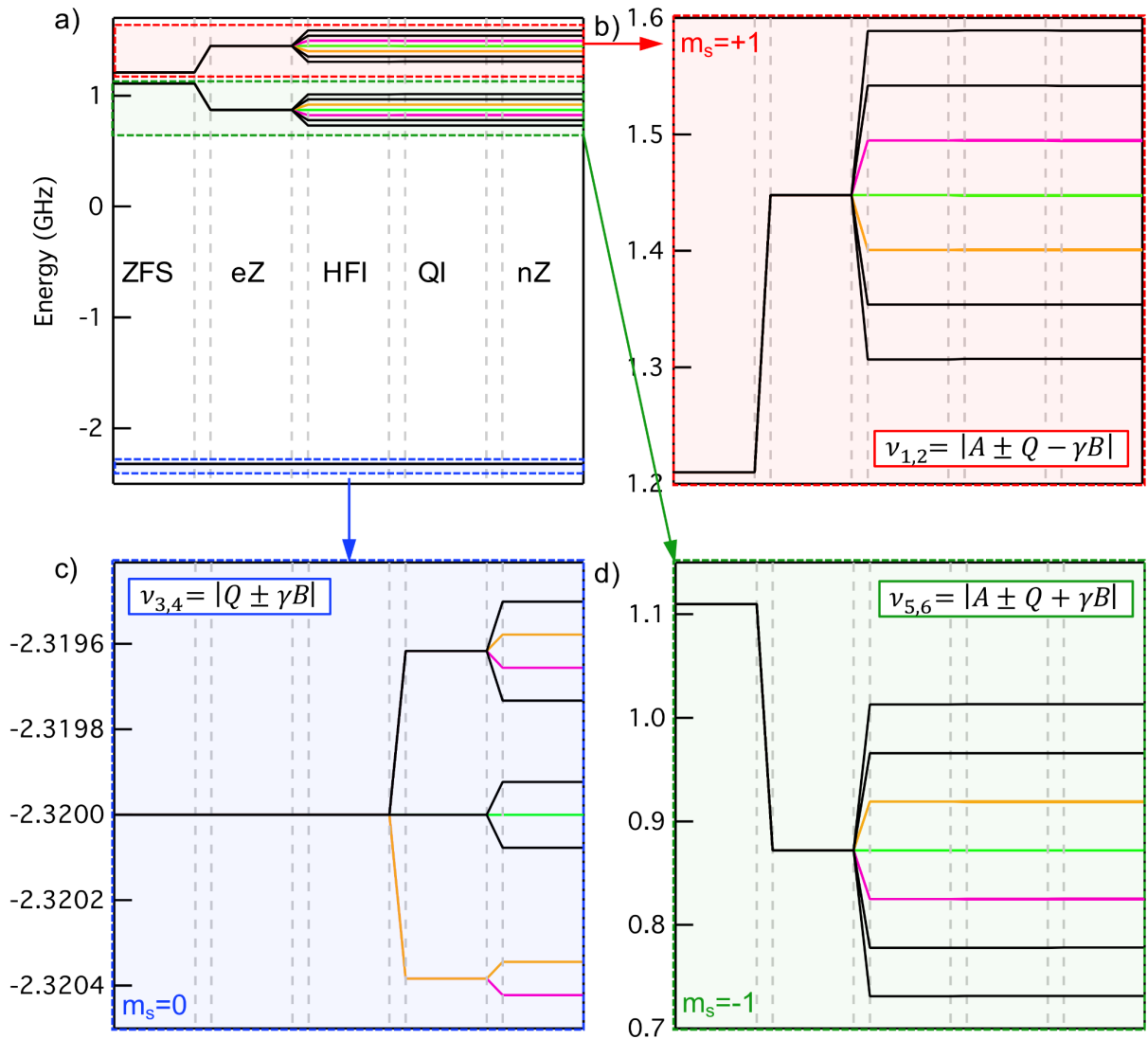

**Supporting Figure S1.** Energy diagram corresponding to spin-Hamiltonian (Eq. (1)) with 6 possible ESEEM frequencies taking three equivalent $^{10}$N nuclei into account (B=8.5 mT). a) Overview of all energy levels. b)-d) Zoom into the energy levels corresponding to the different $m_s$ states. The pink, green and orange energies are representing the participating sub-levels which were addressed in the hole burning measurements (see Fig 4, same colors). Within the energy levels of the different $m_s$ states, we derive 6 possible ESEEM frequencies. Since the hyperfine coupling A=47 MHz contributes to $\nu_{1,2}$ and $\nu_{5,6}$, they are not observable for the mentioned ESEEM measurements in the 100 kHz regime leaving $\nu_{3,4}$ as the detectable frequencies, as discussed in the main text.

**Three-pulse ESEEM**

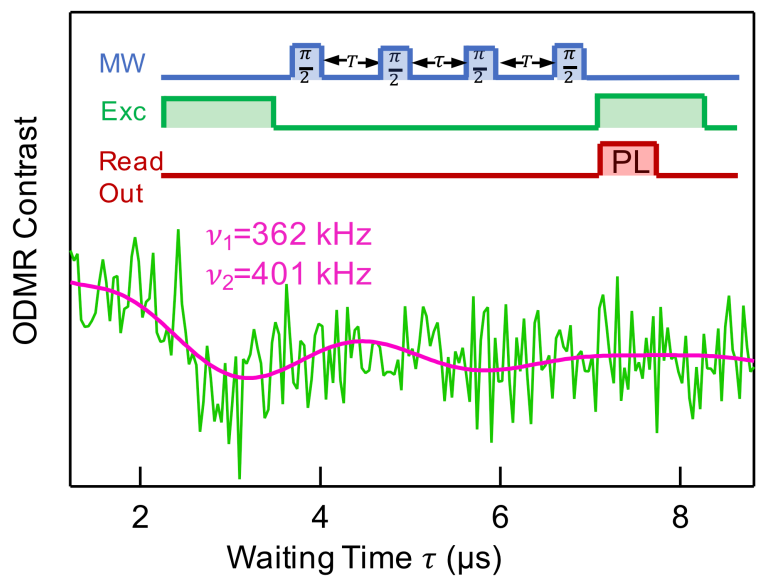

**Supporting Figure S2.** *Three-pulse ESEEM at 4 mT: In order to confirm the ESEEM frequencies extracted from the $T_2$ measurements a three-pulse ESEEM measurement is applied to the system. The corresponding pulse sequence for the microwave pulses is depicted in blue, the laser pulses in green and the detection of the PL in red. In contrast to three-pulse ESEEM in pulsed EPR measurements, a fourth pulse is required at the position of the expected echo in order to return the spin system into a well-defined state, which enables the optical readout. The resulting curve is then limited by T1 instead of T2, which can be fitted with two frequencies.*